\documentclass[usenatbib]{mn2e}
\usepackage{epsfig,rotating,natbib,amssymb,amsmath,times}

\title[The impact of magnetic fields on single and binary star formation]{The impact of magnetic fields on single and binary star formation}

\author[Price \& Bate]{Daniel J. Price and Matthew R. Bate \\
School of Physics, University of Exeter, Stocker Rd, Exeter EX4 4QL, UK \\
}
\date{Submitted: 24th November 2006 Revised: 19th January 2007}
\pagerange{\pageref{firstpage}--\pageref{lastpage}} \pubyear{2006}

\begin{document}
\label{firstpage}
\bibliographystyle{mn2e}
\maketitle

\begin{abstract}
We have performed magnetohydrodynamic (MHD) simulations of the collapse and fragmentation of molecular cloud cores using a new algorithm for MHD within the smoothed particle hydrodynamics (SPH) method, that enforces the zero magnetic divergence constraint.  We find that the support provided by magnetic fields over thermal pressure alone has several important effects on fragmentation and the formation of binary and multiple systems, and on the properties of massive circumstellar discs.  The extra support suppresses the tendency of molecular cloud cores to fragment due to either initial density perturbations or disc fragmentation.  Furthermore, unlike most previous studies, we find that magnetic pressure plays the dominant role in inhibiting fragmentation rather than magnetic tension or magnetic braking.  In particular, we find that if the magnetic field is aligned with the rotation axis of the molecular cloud core, the effects of the magnetic field on fragmentation and disc structure are almost entirely due to magnetic pressure, while if the rotation axis is initially perpendicular to the magnetic field, magnetic tension plays a greater role and can actually aid fragmentation.  Despite these effects, and contrary to several past studies, we find that strongly-perturbed molecular cloud cores are able to fragment to form wide binary systems even in the presence of quite strong magnetic fields.  For massive circumstellar discs, we find that slowing of the collapse caused by the magnetic support decreases the mass infall rate on to the disc and, thus, weakens gravitational instabilities in young massive circumstellar discs.  This not only reduces the likelihood that they will fragment, but also decreases the importance of spiral density waves in providing angular momentum transport and in promoting planet formation.
\end{abstract}


\section{Introduction}
 
The ability of a magnetic field to support a cloud against gravity may be quantified using the \emph{mass-to-flux} ratio, which for a spherical cloud is given by
\begin{equation}
\frac{M}{\Phi} \equiv \frac{M}{4\pi R^{2} B_{0}}.
\end{equation} 
where $M$ is the mass contained within the cloud volume, $\Phi$ is the magnetic flux threading the cloud surface at radius $R$ assuming a uniform magnetic field $B_{0}$. By a straightforward application of the virial theorem it can be shown that there exists a critical value of $M/\Phi$ below which a cloud will be supported against gravitational collapse by the magnetic field. For a uniform, spherical cloud this critical value takes the form \citep[e.g.][]{mestel99,mk04}.
\begin{equation}
\left(\frac{M}{\Phi}\right)_{\rm crit} = \frac{2 c_{1}}{3} \sqrt{\frac{5}{\pi G \mu_{0} }},
\label{eq:mphicrit}
\end{equation}
where $G$ and $\mu_{0}$ are the gravitational constant and the permeability of free space respectively and $c_{1}$ is a parameter determined numerically by \citet{ms76} to be $c_{1} \simeq 0.53$.

Observations of magnetic fields in molecular clouds (\citealt{crutcher99}; see \citealt{hc05} for a recent review) show that magnetic fields are routinely observed with strengths where magnetic and gravitational forces are comparable (i.e. the mass-to-flux ratio is close to the critical value).  Thus, magnetic effects must be included in a complete theory of star formation.  Despite this fact, most three-dimensional numerical simulations of star formation have neglected the effects of magnetic fields.  To date, only a few works have begun to examine their impact on the star-formation process.

The earliest three-dimensional magnetohydrodynamic (MHD) simulations of molecular cloud cores were those performed by \citet{dorfi82} using a Cartesian grid code and \citet{benz84} using a smoothed particle hydrodynamics (SPH) code.  Both studies found that magnetic fields inhibit the collapse of molecular cloud cores and could lead to the formation of bar-shaped structures during collapse.  Dorfi also noted that magnetic braking can remove the bulk of a core's angular momentum over a period of time that is short compare with the core's free-fall time.  Indeed, magnetic fields are often invoked to solve the `angular momentum problem' in star formation by braking the rotation of molecular cloud cores.

\citet{phillips86a,phillips86b} aimed to investigate whether magnetic fields enhanced or inhibited fragmentation of collapsing molecular cloud cores using an early MHD SPH code.  He found that none of his magnetised clouds fragmented.  However, given that none of his unmagnetised clouds fragmented either, it is likely his results suffered from a lack of spatial resolution due to the use of a large, spatially-uniform smoothing length.

More recently, Boss performed hydrodynamical simulations of the collapse of molecular cloud cores where the effect of magnetic fields was approximated by increasing the thermal pressure to mimic magnetic pressure and modelling magnetic tension forces as a dilution of self-gravity \citep{boss00,boss02}.  He found that approximating magnetic tension effects as a dilution of self-gravity lead to fragmentation being enhanced.  However, these calculations did not model the full MHD equations which may impact the applicability of his conclusions. 

\citet{hw04b} returned to the problem of the impact of magnetic fields on fragmentation using an SPH code with variable smoothing lengths (spatial resolution).  They used similar MHD equations to those of Phillips, but in addition they used a two-fluid (ion/neutral) approach that allowed them to model partially ionised fluids and take into account the effects of ambipolar diffusion.  They began with a rotating molecular cloud core that collapsed to form a central object surrounded by a disc which fragmented to form companion objects when the simulation was performed without magnetic fields.  They then investigated the effect of magnetic fields.  Beginning with a field strong enough to make the initial conditions subcritical, the cloud evolved via ambipolar diffusion until it became supercritical and collapsed.  However, they found that angular momentum was transported outward due to magnetic braking at such a rate that fragmentation did not occur.  Indeed, in most of their cases, a disc did not even form around the central object; the coupling between the ions and neutrals had to be reduced by a factor of 25 even to obtain a small disc.  Thus, they concluded that magnetic fields inhibit fragmentation.  However, they were also unable to follow the calculations for long periods of time because the divergence of the magnetic field was not constrained to be zero and the calculation had to be terminated when it grew too large.  \citet{ziegler05} and \citet{fht06} each performed similar calculations to Hosking \& Whitworth, but under the ideal MHD approximation, to test new versions of the NIRVANA finite-difference code and the RAMSES Godunov adaptive mesh refinement code, respectively.  They obtained very similar results to those of Hosking \& Whitworth both with and without magnetic fields (i.e., fragmentation without magnetic fields and the formation of a single object when magnetic fields were included).

\citet{machida04,machida05a,machida05b}  followed the collapse of a wide range of initially magnetised rotating molecular cloud cores using a three-dimensional grid-based MHD code.  They found that magnetic fields suppressed fragmentation, but presenting their results in a rotation versus magnetic field plot they found that fragmentation still occurred if the cloud rotation rate was increased along with the magnetic field strength (i.e., for a given field strength, there was a rotation rate below which the cloud did not fragment and for stronger fields this critical rotation rate increased).  Fragmentation in their calculations occurred via either ring fragmentation (weaker field strengths) or bar fragmentation (stronger field strengths).  

Finally, three-dimensional MHD simulations have also begun to be employed to study protostellar jet formation \citep{mt04,machida05b,machida06,bp06}.  In particular, Banerjee \& Pudritz performed calculations similar to those we present here.  They began with a rotating dense core embedded in a static hot low-density medium with a weak magnetic field (ratio of thermal pressure to magnetic pressure of 84) parallel to the rotation axis of the core.  They found that the rotating core lost angular momentum due to the launching of torsional Alfven waves into the low-density medium.  The simulations began with a rotation rate that, in the absence of a magnetic field, resulted in the formation a ring that fragmented into a binary.  With a magnetic field, the size of the ring was reduced dramatically (by approximately two orders of magnitude) due to angular momentum transport.  As the simulation progressed, they observed the outward propagation of a large magnetically-driven bubble above and below the midplane and, later, the formation of a sub-AU scale jet from the central object.

In summary, the calculations discussed above clearly confirm that magnetic fields act to transport angular momentum within rotating cores and to external media.  Most past calculations also indicate that magnetic fields suppress the formation of binary and multiple systems.  However, it is still not yet clear what level of magnetic field is required to inhibit binary formation and the dependence of this critical field strength on the initial cloud parameters (e.g. density perturbations, rotation rate, and/or turbulence).  For example, \citet{phillips86a,phillips86b} and \citet{hw04b} find no fragmentation in any of their calculations that included magnetic fields, \citet{bp06} produce a ring that does manage to fragment into a binary system, albeit one with a very small separation of less than 0.1 AU, while \citet{machida05b} are able to form binaries with large separations of $\sim 100$ AU.

In this paper, we report the first results from numerical calculations of the collapse of molecular cloud cores using a recently developed method for Smoothed Particle Magnetohydrodynamics (SPMHD) \citep[][hereafter PM05]{pm05}.  The calculations were performed in the ideal MHD approximation.  Thus, we are limited to studying the collapse of clouds that are initially supercritical (i.e. they collapse without the need for ambipolar diffusion).  It is straightforward to extend the method in the future to model ambipolar diffusion (e.g. by adapting the two-fluid method of \citealt{hw04a}).  However, given the differing conclusions that currently exist in the literature as to the effects of magnetic fields on fragmentation as discussed above, we have chosen to begin with the simplest case first.  Our aims are first, to demonstrate the potential of this new method for star formation applications, and second, to investigate the issue of how magnetic fields impact the evolution and fragmentation of collapsing molecular cloud cores due to the extra support and the angular momentum transport they provide. We note in passing that a parallel but complementary approach to developing an algorithm for MHD in SPH has been undertaken by Steinar B{\o}rve and colleagues \citep{bot01,bot06} based on a regularisation of the underlying particle distribution, although this algorithm has not yet been applied to star formation problems.

The numerical method is discussed in \S\ref{sec:numerics}. The setup of the simulations is discussed in \S\ref{sec:initconds} and results are presented in \S\ref{sec:results}. These results are discussed in \S\ref{sec:discussion} and summarised in \S\ref{sec:summary}.

\section{Numerical method}
\label{sec:numerics}

\subsection{Hydrodynamics}
 The hydrodynamic method is based on a binary tree smoothed particle hydrodynamics (SPH) code originally written by \citet{benzetal90}. Important modifications were made by \citet{bate95} in the form of sink particles (see below) and the use of individual particle timesteps. In developing the code for MHD calculations we have also made several improvements to the hydrodynamics, most significantly in the form of incorporating the so-called `variable smoothing length terms' \citep{sh02,monaghan02,price04} which ensure that both energy and entropy conservation are maintained to timestepping accuracy. The smoothing length for each particle is spatially adapted with the density according to the rule
\begin{equation}
h = \eta \left( \frac{m}{\rho} \right)^{1/3},
\end{equation}
where $h$ is the smoothing length, $m$ is the particle mass, $\rho$ is the density and $\eta$ is a dimensionless parameter which we set to $h=1.2$, corresponding to an average of $\sim 60$ neighbours for each particle. In the variable smoothing length formulation of \citet{sh02} this is a non-linear equation for both $h$ and $\rho$ (that is, $\rho_{i}$ for particle $i$ is calculated via an SPH summation which depends on $h_{i}$) that we solve iteratively \citep{pm06}. The SPH equations are integrated using a standard second-order leapfrog method with individual timesteps for each particle \citep{bbp95}.
 
  Gravitational force softening is performed using the usual SPH cubic spline kernel with an adaptive softening length set equal to the local smoothing length. We use the formalism developed recently by \citet{pm06} which ensures that momentum and energy conservation are retained even in the presence of a spatially variable softening length.
 
  Artificial viscosity and, for MHD, resistivity are applied as in PM05 which in the case of viscosity, is equivalent to the formalism given by \citet{monaghan97} with the dimensionless parameters controlling the artificial viscosity and resistivity allowed to evolve with time as in \citet{mm97} (generalised for resistivity in PM05). It should be noted that the \citet{mm97} artificial viscosity switch is not particularly useful during the collapse phase as it simply responds to the velocity divergence due to the gravitational infall (rather than shocks). Whilst alternative switches could be investigated, here we simply cap the maximum value for the viscosity parameter to 1.1.

\subsection{Equation of state}
  We use a barotropic equation of state for the thermal pressure of the form
\begin{equation}
P = K \rho^{\gamma}.
\end{equation}
The polytropic exponent $\gamma$ changes according to
\begin{eqnarray}
\gamma = 1,  & & \rho \le 10^{-14} {\rm g\phantom{l}cm}^{-3}, \nonumber \\
\gamma = 7/5, & &  \rho > 10^{-14} {\rm g\phantom{l}cm}^{-3}, \nonumber
\label{eq:eos}
\end{eqnarray}
giving an isothermal equation of state for low densities but a transition at higher densities which reflects the heating of the gas as it becomes optically thick. The value of $K$ is set equal to the square of the isothermal sound speed $c_{\rm s}$ at low densities and changes at the transition density so as to make the pressure a continuous function of density. We set the transition of the equation of state at a comparatively low density ($10^{-14}$g cm$^{-3}$) in order to inhibit disc fragmentation in the calculations, since in this paper we are interested in the effects of magnetic fields on fragmentation that is seeded by initial density perturbations rather than on whether the discs that form in the system subsequently fragment to produce a higher-order multiple system.

\subsection{Sink particles}
Sink particles were introduced by \citet{bbp95} in order to follow star formation calculations beyond the initial collapse phase in an efficient manner. When pressure-supported fragments form in the calculations due to the parametrized equation of state discussed above, the high densities and short dynamical times means that it becomes computationally expensive to follow the internal evolution of these protostars over timescales comparable to the initial free-fall time of the cloud. Instead, for the calculations presented here, a sink particle is inserted once the peak density exceeds $\rho_{\rm s} = 10^{-10}$g cm$^{-3}$. 
 
 The sink particle is formed by replacing the SPH gas particles contained within $r_{\rm acc} = 6.7$AU of the densest gas particle in a pressure-supported fragment by a point mass with the same mass and momentum. Gas which later falls onto this point mass is accreted if it is bound and has a specific angular momentum which is lower than that required to form a circular orbit at the radius $r_{\rm acc}$ from the sink particle. Sink particles interact with the gas only via gravity and accretion. When magnetic fields are present, this means that the magnetic field in the central regions is removed by the sink particle and consequently not allowed to feed back on the surrounding cloud. Whilst this is obviously a somewhat crude approximation at present, it enables us to study the subsequent accretion of gas to form a circumstellar disc and, for calculations that form binaries, the subsequent evolution of the binary system without the calculation grinding to a halt trying to follow the internal dynamics of the protostars themselves.

\subsection{Magnetohydrodynamics}
 The magnetohydrodynamics implemented is an extension of the methods for MHD in SPH developed recently by \citet{pm04a,pm04b} and PM05. The extension is that here, as in \citet{pr06}, the divergence constraint on the magnetic field is satisfied by expressing the magnetic field as a function of two scalar variables $\alpha$ and $\beta$ according to
\begin{equation}
{\bf B} = \nabla\alpha \times \nabla \beta,
\label{eq:eulerpots}
\end{equation}
 which we refer to as the `Euler potentials' \citep{stern70} but elsewhere referred to as the `Clebsch formalism' \citep{pm85} or `flux co-ordinates'. We calculate the gradients of the potentials using an SPH summation which is exact for a linear gradient \citep{price04}, that is,
\begin{eqnarray}
\chi_{\mu\nu} \nabla^{\mu} \alpha_{i} & = & -\sum_{j} m_{j} (\alpha_{i} - \alpha_{j}) \nabla^{\nu}_{i}W_{ij}(h_{i}), \label{eq:dalpha} \\
\chi_{\mu\nu} \nabla^{\mu} \beta_{i} & = & -\sum_{j} m_{j} (\beta_{i} - \beta_{j}) \nabla^{\nu}_{i}W_{ij}(h_{i}), \label{eq:dbeta}
\label{eq:spheuler}
\end{eqnarray}
where latin indices refer to the particles, greek indices refer to vector components and the matrix quantity $\chi_{\mu\nu}$ is given by
\begin{equation}
\chi_{\mu\nu} = \sum_{j} m_{j} (r_{i}^{\mu} - r_{j}^{\mu}) \nabla^{\nu} W_{ij} (h_{i}). \label{eq:chi}
\end{equation}
The gradient calculations involve solving a $3\times 3$ matrix equation in each case, but this is straightforward to solve analytically and implementation in the code requires only the temporary storage of the matrix variable $\chi$ for a given particle (ie. we do not need to store calculated values for more than one particle). Furthermore, since the summations (\ref{eq:dalpha}), (\ref{eq:dbeta}) and (\ref{eq:chi}) do not involve the density on neighbouring particles, the gradients can be calculated efficiently alongside the usual density summation. The $B$ field calculated from the Euler potential gradients in this manner is then used in the force equation as in PM05.
 
  Apart from satisfying the divergence constraint, the other main advantage of using the Euler potentials representation is that, for ideal MHD, they evolve according to
\begin{equation}
\frac{d\alpha}{dt} = 0, \hspace{1cm} \frac{d\beta}{dt} = 0,
\end{equation}
corresponding to the advection of magnetic field lines by Lagrangian particles \citep{stern66}. We extend the Euler potentials method to non-ideal MHD by incorporating shock-capturing dissipation terms, the form of which is given, using a simple generalisation of the terms derived in \citet{pm04b}, by
\begin{eqnarray}
\left(\frac{d\alpha}{dt}\right) _{\rm diss} & = & \sum_{j} m_{j} \frac{\bar{\alpha}^{\rm B}_{ij} v_{sig}}{\bar{\rho}_{ij}} (\alpha_{i} - \alpha_{j}) \vert \overline{\nabla W_{ij}} \vert,\label{eq:euler1diss} \\
\left(\frac{d\beta}{dt}\right) _{\rm diss} & = & \sum_{j} m_{j} \frac{\bar{\alpha}^{\rm B}_{ij} v_{sig}}{\bar{\rho}_{ij}} (\beta_{i} - \beta_{j}) \vert \overline{\nabla W_{ij}} \vert \label{eq:euler2diss},
\end{eqnarray}
where the summation is over neighbouring particles, $v_{\rm sig}$ is a maximum signal velocity between the particle pair as in PM05, the mean density $\bar{\rho} = 0.5(\rho_i + \rho_{j})$, $\vert \overline{\nabla W_{ij}} \vert$ refers to the magnitude of the mean kernel gradient  $\overline{\nabla W_{ij}} = 0.5[\nabla W_{ij}(h_{i}) + \nabla W_{ij}(h_{j})]$ and $\alpha^{\rm B}$ is a time-variable co-efficient for each particle that is evolved as described in PM05.

The magnetic force is computed using the \citet{morris96} formalism discussed in PM05 (using the ${\bf B}$ computed from equation \ref{eq:eulerpots}) which ensures stability of the SPMHD formalism against particle-clumping instabilities in the regime where gas pressure is dominant over magnetic pressure. The MHD part of the force equation (that is, apart from the gravitational and artificial viscosity forces) reads
\begin{eqnarray}
&& \frac{dv_{i}^{\mu}}{dt} = \nonumber \\
&& - \sum_j m_j\left[\frac{P_i + \frac{1}{2\mu_{0}} B^2_i}{\Omega_{i} \rho_i^2}\nabla^{\mu} W_{ij} (h_{i}) + 
\frac{P_j + \frac{1}{2\mu_{0}}
B^2_j}{\Omega_{j}\rho_j^2}\nabla^{\mu} W_{ij} (h_{j})\right]\nonumber \\
&& +  \sum_j m_j
\frac{(B^\mu B^\nu)_i - (B^\mu B^\nu)_j}{\mu_{0}\rho_i\rho_j}\overline{\nabla_{\nu} W_{ij}}, 
\label{eq:morrisforce}
\end{eqnarray}
where $v_{i}, \rho_{i}, P_{i}$ and $B_{i}$ refer to the density, pressure and magnetic field of particle $i$ and $\Omega$ is a normalisation term related to the gradient of the smoothing length (as in, e.g. \citealt{monaghan02,pm06}). The first term in (\ref{eq:morrisforce}) is the isotropic hydrodynamic + magnetic pressure force and the second term is the magnetic tension force. For simulations ``without magnetic tension'' we do not include the latter term.

 A detailed summary of the recent changes to the hydrodynamic method (including details of the implementation of the variable smoothing length SPH formalisms in both the pressure and gravity terms) and the implementation of the Euler potentials into the numerical code (including test problems comparing the use of them to the `standard' SPMHD formalism of PM05) are discussed in detail in \citet{pr07} and we refer the reader to this paper for an up-to-date summary of the present numerical code (the specific code described differs from that used here but the algorithms implemented in each are identical).

\section{Initial conditions}
\label{sec:initconds}
 The initial cloud is a sphere of radius $R= 4\times 10^{16}$ cm (0.013 pc) and mass $M= 1M_{\odot}$ with mean density $\rho_{0} = 7.43\times 10^{-18}$g~cm$^{-3}$. The free-fall time of the cloud is $t_{\rm ff} = 2.4\times 10^{4}$ years.
 
 We assume, for simplicity, that an initially uniform magnetic flux threads the cloud and connects it to the surrounding interstellar medium. However, a key factor in the problems studied here is the angular momentum transfer introduced by the magnetic field in the form of magnetic braking of the rotating core. Thus careful attention must be paid to the boundary condition at $r=R$. Experiments with simple boundary conditions for SPH (for example, using constant pressure boundaries or ghost particles) proved somewhat unsatisfactory, particularly because, in the higher magnetic field strength runs, significant material is flung outwards by the cloud along the magnetic field lines into the surrounding medium. We therefore model the boundaries self-consistently by placing the cloud within a uniform, low density box of surrounding material in pressure equilibrium with the cloud \citep[see also ][]{hosking02,bp06}. To ensure regularity of the particle distributions at the box boundaries, we use quasi-periodic boundary conditions at the box edge (that is, particles within $2$ smoothing lengths of the boundary are `ghosted' to the opposite boundary, with no self gravity between SPH particles and ghost particles). Since a uniform magnetic field necessarily implies a linear gradient in the Euler potentials, continuity of the magnetic field across the box boundary is ensured by adding an offset to the values of $\alpha$ and $\beta$ copied to the ghosted particles corresponding to an extrapolation of the linear gradients outside the box boundaries.
 
\begin{figure}
\begin{center}
\begin{turn}{270}\epsfig{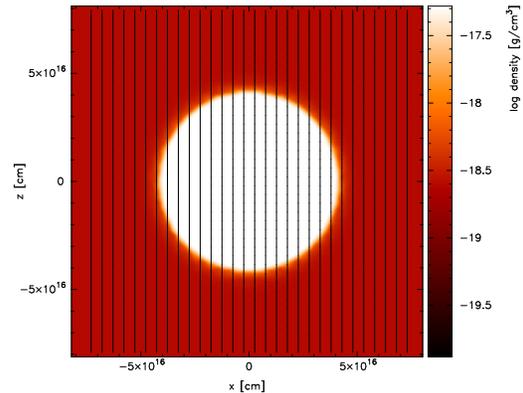}\end{turn}
\caption{Initial setup. A uniform density cloud of radius $4\times10^{16}$cm and mass 1$M_{\odot}$ in uniform rotation is placed in a warm, low density surrounding medium (density ratio 30:1). Magnetic field lines thread the cloud. For the binary calculations an $m=2$ perturbation is applied to the cloud density. The figure shows a cross section slice of the density at $y=0$ together with magnetic field lines aligned with the rotation (z) axis.}
\label{fig:setup}
\end{center}
\end{figure}

  We find that satisfactory results are obtained using a box size of $-8 \times 10^{16}$cm $< x,y,z < 8 \times 10^{16} $cm (that is, twice the cloud radius in each direction) and a density ratio of $30:1$ between the cloud and the surrounding medium. This density ratio was chosen simply to ensure that the surrounding medium is sufficiently hot so as not to contribute significantly to the self-gravity of the cloud (that is, $c_{\rm s}^{2} > GM/R$). The initial setup is shown in Figure~\ref{fig:setup}, showing a cross-section slice of density at $y=0$ with overlaid magnetic field lines for a field initially oriented in the $z-$direction.

  Both the spherical cloud and the surrounding medium are set up by placing the particles in a regular close-packed lattice arrangement \citep[e.g.][]{hosking02} which is a stable arrangement for the particles \citep{morrisphd}. Whilst such regularity introduces some undesirable side effects due to the lattice regularity in the initial collapse phase, these small and transient effects are largely eliminated by the time star formation occurs. The exact positions of the box boundaries are adjusted slightly to ensure continuity of the lattice across the ghosted boundary.
  
  The magnetic field strength is characterised in terms of a mass-to-flux ratio expressed in units of the critical value (\ref{eq:mphicrit}). The corresponding magnetic field strength is given by
\begin{equation}
B_{0} = 814\mu G \left(\frac{M}{\Phi} \right)^{-1} \left(\frac{M}{1 M_{\odot}}\right) \left(\frac{R}{0.013 {\rm pc}}\right)^{-2},
\end{equation}
where $M/\Phi$ is the mass to flux ratio in units of the critical value.

 For both the axisymmetric and binary star formation problems, we calculate a sequence of collapse models with magnetic field strengths corresponding to mass-to-flux ratios, in units of the critical value (\ref{eq:mphicrit}), of $\infty$ (that is, hydrodynamic), $100, 20, 10, 7.5, 5, 4, 3, 2$ and $1$. For our choice of cloud mass and radius these correspond to field strengths of $B=0, 8.1, 40.7, 81.3, 108.5, 163, 203, 271, 407$ and $814 \mu G$ respectively. The initial Alfv\'en speeds in the cloud are therefore $0, 8.4\times 10^{-2}, 0.42, 0.84, 1.12, 1.68, 2.10, 2.80, 4.21$ and $8.42$ km/s respectively. We compute these sequences in each case both for a magnetic field initially aligned with the rotation axis (ie. oriented in the $z-$direction) and for a magnetic field initially oriented in the $x-$direction, thus bracketing two geometric extremes. We find that, whilst there are similar global trends in both cases due to the effect of magnetic pressure, the orientation of the magnetic field with respect to the rotation axis plays an important role in determining the outcome.
 
 
  Simulations are performed using 300,000 particles in the cloud itself. Including the external medium as discussed above, this results in a total of 451,233 particles in each simulation. We have also performed simulations using 30,000 and 100,000 particles in the cloud, which show only minor differences in results (see discussion in \S\ref{sec:binary}). In order to resolve the local Jeans mass throughout the calculations, we require at least 30,000 particles for our chosen equation of state \citep{bateburkert97}.  Thus, all our calculations resolve the local Jean mass, and the high resolution calculations presented here do so by an order of magnitude in particle number.
 
\section{Results}
\label{sec:results}

\subsection{Axisymmetric collapse}
\label{sec:axisym}
For an axisymmetric collapse, we set the initial cloud to be uniform density in uniform rotation with an angular velocity of $\Omega = 1.77 \times 10^{-13}$ rad s$^{-1}$ corresponding to a ratio of rotational to gravitational energy $\beta_{\rm r} \simeq 0.005$ and $\Omega t_{\rm ff} = 0.136$.

  The initial cloud temperature is set such that the ratio of thermal to gravitational energy $\alpha = 0.35$ (where by gravitational energy we mean the magnitude of the gravitational potential energy). This corresponds to an internal energy of $7.04\times 10^{8}$ergs~g$^{-1}$ and, assuming a mean molecular weight for molecular hydrogen (ie. $\mu = 2$), corresponds to an isothermal temperature of $11.3$K. The initial sound speed in the cloud is $c_{\rm s} = 2.16\times 10^{4}$~cm~s$^{-1}$ (for comparison the sound speed in the external medium is $c_{\rm s,medium} = 11.9\times 10^{4}$~cm~s$^{-1}$, ie. $c_{\rm s,medium}^{2} \simeq 4.2 GM/R$).

Expressed in terms of the ratio of gas to magnetic pressure, $\beta_{\rm m}$, our chosen sequence of mass to flux ratios of $\infty, 100, 20, 10, 7.5, 5, 4, 3, 2$ and $1$ in this cloud correspond to $\beta_{\rm m} = \infty, 983, 39, 9.8, 5.5,  2.5, 1.6, 0.85, 0.39$ and $0.098$, respectively. 

 We compute these sequences for initial magnetic fields threading the cloud which are aligned parallel and perpendicular to the rotation axis (that is, in the $z-$ and $x-$ directions respectively in our computational domain).

\subsubsection{Initial field aligned with the rotation axis}
\label{sec:axisymBz}

\begin{figure}
\begin{center}
\begin{turn}{270}\epsfig{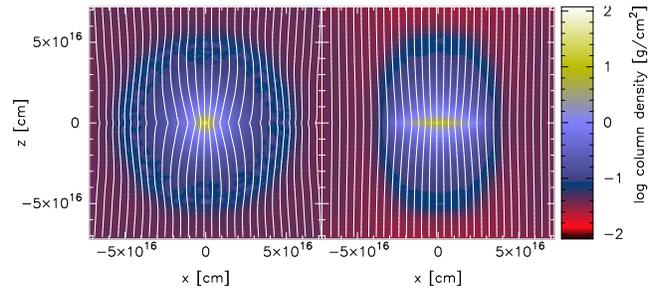}\end{turn}
\caption{Column density and projected magnetic field lines in the collapsing, axisymmetric cloud, showing results at $t_{\rm ff}=1.01$ using mass-to-flux ratios in units of the critical value of (left) 100 (that is, a very weak field) and (right) 3 (ie., a very strong field) with an initial field aligned with the rotation (z) axis. In the weak field case (left) the magnetic field lines are deformed by the cloud, leading to an almost-spherical collapse, whereas for strong fields (right) material is strongly channelled along magnetic field lines, leading to an bipolar-shaped cloud and a pancake-like collapse.}
\label{fig:fieldlines}
\end{center}
\end{figure}

 As an illustration of the global cloud evolution for runs with the field aligned with the rotation axis, Figure~\ref{fig:fieldlines} shows the column density and projected magnetic field lines in the collapsing cloud at $t=1.01$ free fall times for two runs with a very weak ($M/\Phi = 100$, left panel) and very strong ($M/\Phi = 3$, right panel) magnetic field, projected in the $z-x$ direction (integrated through the $y-$direction). For runs with weak magnetic fields ($M/\Phi \gtrsim 10$), the collapse is almost spherical and the magnetic field lines are bent by the infalling gas flow, developing a toroidal component of similar magnitude to the $B_{\rm z}$ field. For very strong magnetic fields ($M/\Phi \lesssim 3$) the collapse is strongly channelled along the magnetic field lines, producing a pancake-like collapse and developing only a small toroidal component in the large scale magnetic field. In the latter case the interaction between the strong magnetic field and the rotating cloud is also evident by the shape of the cloud in the right hand panel of Figure~\ref{fig:fieldlines}, which has evolved to a bipolar configuration as material is flung outwards along the magnetic field lines.
 
\begin{figure*}
\begin{center}
\epsfig{file=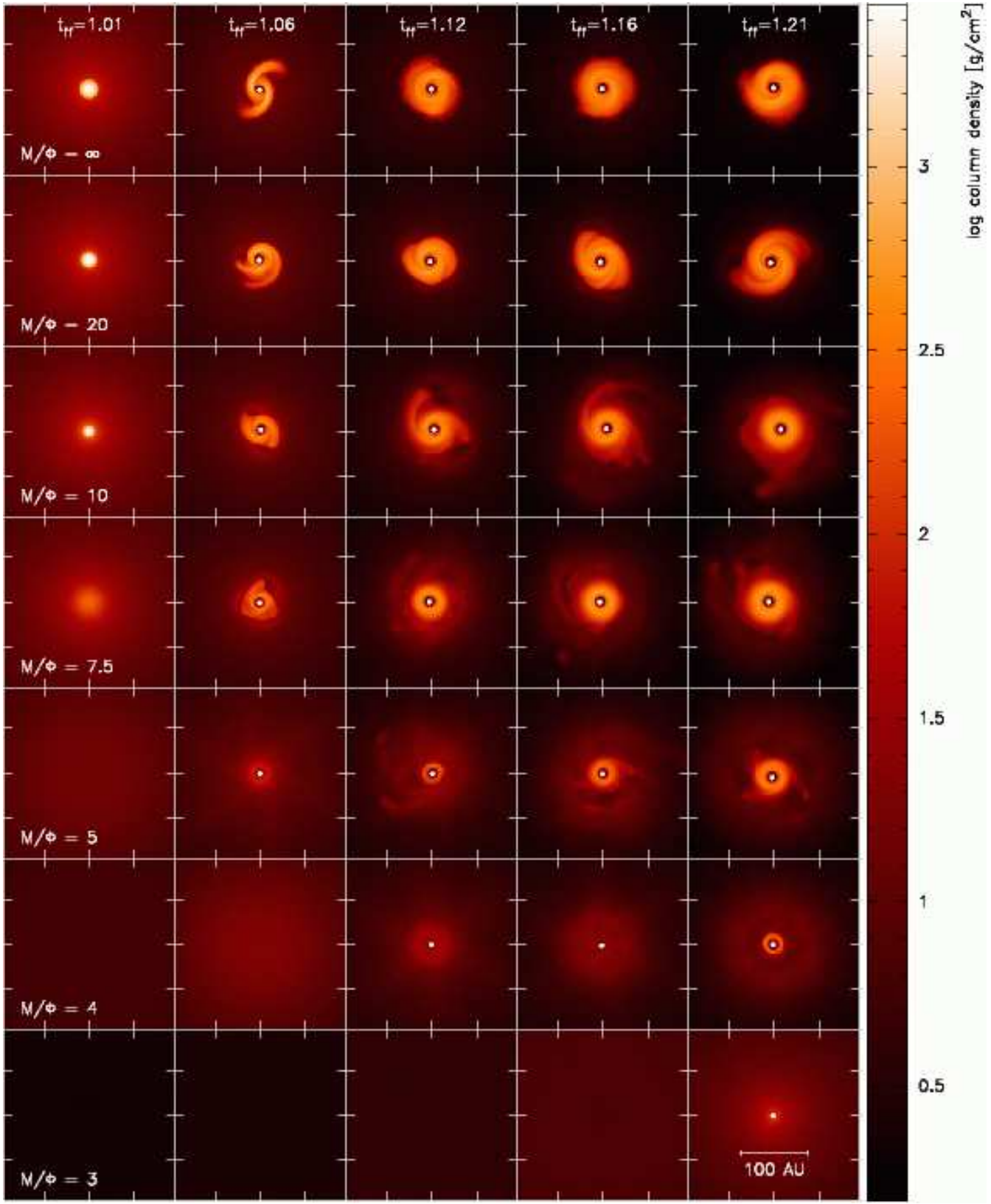,width=\textwidth}
\caption{Results of the axisymmetric collapse calculations with initial magnetic field aligned with the rotation (z) axis, showing column density in the collapsed cloud (integrated through the $z$ direction). Columns from left to right show snapshots at a given time (given in units of $t_{\rm ff}=2.4\times 10^{4}$ yrs), whilst from top to bottom rows show the results for increasing magnetic field strengths, given as a mass to flux ratio in units of the critical value but which correspond to $B=0, 40.7, 81.3, 108.5, 163$ and $203\mu G$ respectively. Increasing the magnetic field strength tends to delay and also suppress disc formation.}
\label{fig:axisymcoldens}
\end{center}
\end{figure*}
 
  The magnetic field strength in the collapsing cloud is found to scale with density approximately as $B \propto \rho^{0.6}$, with a pronounced flattening off in the disc itself for the higher field strength runs. This is in good agreement with a similar result found by \citet{bp06} who considered only a collapse with a relatively low field strength, as here with the initial field aligned with the rotation axis.

 The results from axisymmetric collapse calculations using an initial field aligned with the rotation axis are shown in Figure~\ref{fig:axisymcoldens}, showing column density through the cloud (integrated through the $z-$direction) at 5 different times (left to right) for a sequence of runs of increasing magnetic field strength (top to bottom), where time is shown in units of the free-fall time and magnetic field strengths are expressed in terms of the mass-to-flux ratio in units of the critical value. The figure shows the runs with $M/\Phi = \infty, 20, 10, 7.5, 5, 4$ and $3$. The $M/\Phi = 100$ run differs only slightly from the hydrodynamics case and has not therefore been plotted. For very strong magnetic fields ($M/\Phi < 3$) the collapse is strongly inhibited as the mass-to-flux ratio is close to the critical value.  Gas is strongly channelled along the magnetic field lines (Figure~\ref{fig:fieldlines}) and we find that a ring is formed at the cloud radius (ie. far away from the central regions, and long before any collapse has occurred in the centre) in the mid-plane which becomes gravitationally unstable and fragments.  Because this fragmentation is essentially a boundary effect, we do not discuss these calculations any further.

\begin{figure}
\begin{center}
\begin{turn}{270}\epsfig{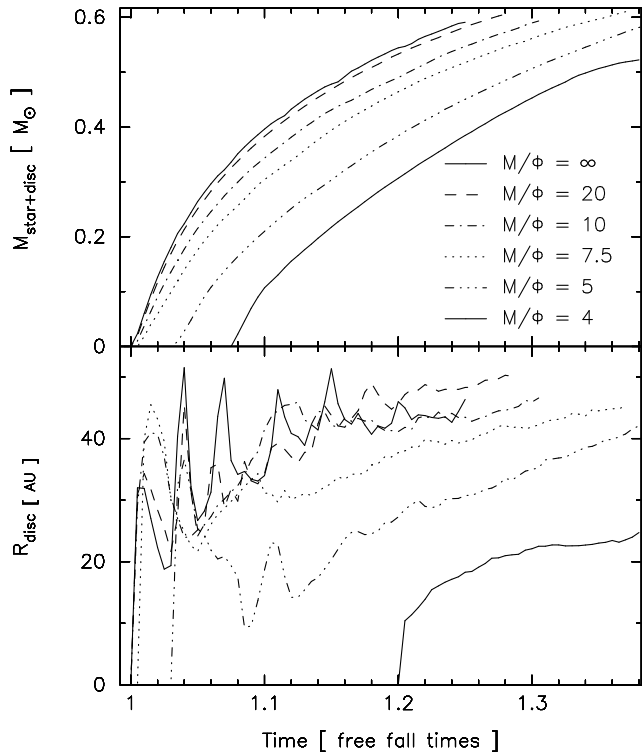}\end{turn}
\caption{Circumstellar disc mass (top) and radius (bottom) plotted as a function of time in the axisymmetric calculations using a field aligned with the rotation axis (see Figure~\ref{fig:axisymcoldens}). The disc mass is defined simply as the mass of material which has a density $> 10^{-13}$g/cm$^{3}$ (ie. an order of magnitude above where the gas becomes optically thick) and the disc radius is defined as the radius containing 99\% of the mass of the star+disc system. For very low magnetic field strengths the disc is more massive and thus dominated by gravitational instabilities resulting in spiral arms, producing a disc radius which is oscillatory in time. Increasing the magnetic field strength tends to suppress disc formation, resulting in smaller, less massive discs which are less dominated by self-gravity.}
\label{fig:discmass}
\end{center}
\end{figure}

The results for intermediate field strengths show a clear trend in both the formation of the protostar and the subsequent size of the disc which forms (Figure~\ref{fig:axisymcoldens}), namely that protostar formation occurs progressively later as the field strength is increased and the disc which forms is smaller, less massive and thus also less prone to gravitational instabilities such as spiral arms which are clearly evident in the hydrodynamic run (top row).

The delay in the onset of fragmentation is largely a result of the extra support provided to the cloud by the magnetic pressure. The size and mass of the resultant disc is affected in this case mainly by the subsequently lower accretion rate onto the central core. Magnetic tension is found to play only a relatively minor role when the field is aligned with the rotation axis. The role of magnetic pressure versus magnetic tension is apparent in the comparison between runs using fields aligned and fields perpendicular to the rotation axis, discussed below.

\begin{figure*}
\begin{center}
\epsfig{file=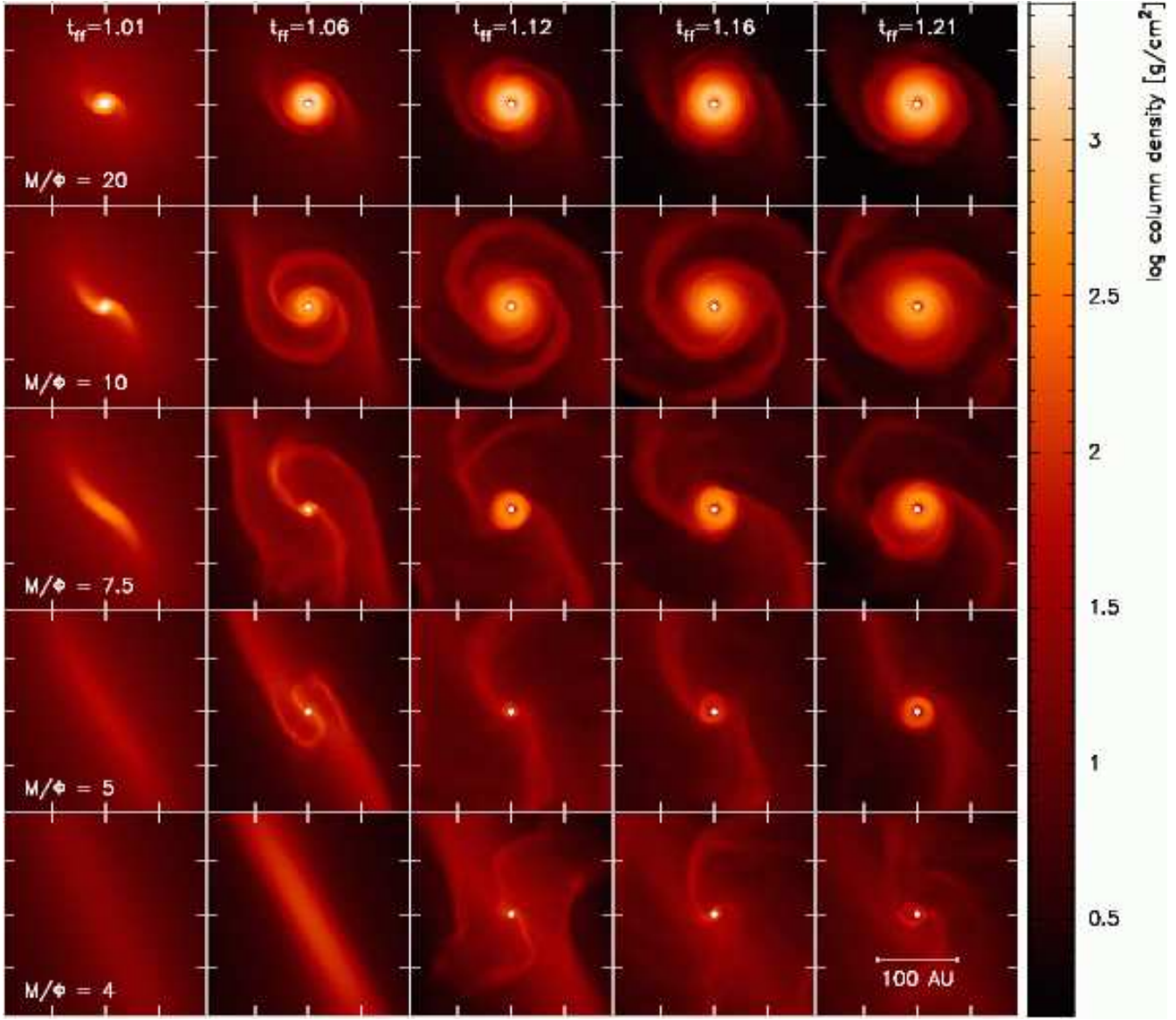,width=\textwidth}
\caption{Results of the axisymmetric collapse calculations with initial magnetic field perpendicular to the rotation axis, which may be compared with the aligned-field runs shown in Figure~\ref{fig:axisymcoldens}. As in the aligned field case, increasing the magnetic field strength tends to delay and also suppress disc formation. In this case, however, the magnetic field geometry results in a increasingly bar-like collapse. For lower field strengths this bar-like structure is wound up to produce low density spiral structure surrounding the disc.}
\label{fig:axisymcoldens_Bx}
\end{center}
\end{figure*}

  We have quantified the results shown in Figure~\ref{fig:axisymcoldens} by calculating the disc mass and radius as a function of time, plotted in Figure~\ref{fig:discmass}. We define the disc mass as simply the mass of material which has a density $> 10^{-13}$g/cm$^{3}$ (ie. an order of magnitude above the density at which the gas has become non-isothermal according to our equation of state \ref{eq:eos}). Similarly the disc radius is defined as the radius which contains 99\% of the mass of the star+disc system. For the lower field strength runs (e.g. the solid line in Figure~\ref{fig:discmass} corresponding to the hydrodynamic run), a disc is formed with a radius of $\sim 30-50$ AU at around 1 free-fall time. The disc radius shows pronounced oscillations in time because the relatively massive discs are dominated by gravitational instabilities which drive non-axisymmetric features such as spiral arms. Increasing the magnetic field strength tends to suppress disc formation, producing smaller, less massive discs with masses and radii which increase more slowly in time and which show a smoother evolution due to the suppression of gravitational instabilities.

 We have also performed axisymmetric calculations with $B_{\rm z}$ fields using higher rotation rates ($\beta_{\rm r} = 0.01$ and $\beta_{\rm r}=0.16$, not shown). Increasing the rotation rate tends to enhance gravitational instabilities in the disc and thus the ability to fragment into multiple systems, however we find that the effect of the magnetic field is the same - namely that increasing the magnetic field strength suppresses gravitational instabilities in the disc and thus the propensity to fragment. 

\subsubsection{Initial field perpendicular to the rotation axis}
\label{sec:axisymBx}

 The results of axisymmetric calculations starting from a magnetic field initially perpendicular to the rotation axis (that is, a field initially in the $x-$direction) are shown in Figure~\ref{fig:axisymcoldens_Bx}, shown at the same times as in Figure~\ref{fig:axisymcoldens} for runs using mass-to-flux ratios of $20, 10, 7.5, 5$ and $4$ (top to bottom), corresponding to the middle 5 rows of Figure~\ref{fig:axisymcoldens}. As in the aligned-field calculations increasing the magnetic field strength can be seen to suppress the formation of the circumstellar disc, forming smaller, less massive discs in which the spiral structure produced by gravitational instabilities is suppressed. However in this case the geometry of the magnetic field results in an increasingly bar-like collapse (most evident in the $M/\Phi = 4$ and $M/\Phi = 5$ runs at $t_{\rm ff} = 1.06$ in Figure~\ref{fig:axisymcoldens_Bx}) which is wound up by the rotation of the core to produce a low density spiral structure surrounding the central disc. The winding up of the bar results in oppositely directed field in the spiral structure which produces significant reconnection, particularly in the stronger field runs where this reconnection results in a small amount of ejected material from the central regions. 
 
  The evolution of disc mass and radius is similar to the aligned-field case (see Figure~\ref{fig:discmass}), mainly because the changes due to magnetic tension are in the lower density material surrounding the disc rather than in the disc itself.  However, we note that the misaligned field seems to inhibit gravitational instabilities in the dense central part of the disc even more than in the aligned-field case.


\subsection{Binary star formation}
\label{sec:binary}
The binary star formation problem we consider is a variation on the `standard isothermal test case' of \citet{bb79}, the main differences in the hydrodynamic problem being the larger cloud radius. The initially uniform density cloud is given a non-axisymmetric $m=2$ density perturbation with amplitude $A$ of the form
\begin{equation}
\rho = \rho_{0}[1 + A\cos{(2\phi)}],
\label{eq:rhoperturb}
\end{equation}
where $\phi$ is the azimuthal angle about the rotation (z) axis and $\rho_{0}$ is set as previously. The cloud is in uniform rotation with $\Omega = 1.006\times 10^{-12}$ rad s$^{-1}$, corresponding to a ratio of rotational energy to gravitational potential energy of $\beta_{\rm r} = 0.16$ and $\Omega t_{\rm ff} = 0.77$. The internal energy of the cloud is set such that the ratio of thermal to gravitational energy is $\alpha = 0.26$. This corresponds to an internal energy of $5.23\times 10^{8}$ergs~g$^{-1}$, a sound speed in the cloud of $1.87\times 10^{4}$cm/s and, assuming a mean molecular weight of $\mu=2$, an isothermal temperature of 8.4K. Given the 30:1 density ratio between the cloud and the surrounding medium and the condition of pressure equilibrium between the two, this means that the sound speed in the external medium is $c_{\rm s,medium} = 10.2 \times 10^{4}$cm/s, and $c_{\rm s,medium}^{2}/(GM/R) = 3.1$.
 
\begin{figure*}
\begin{center}
\epsfig{file=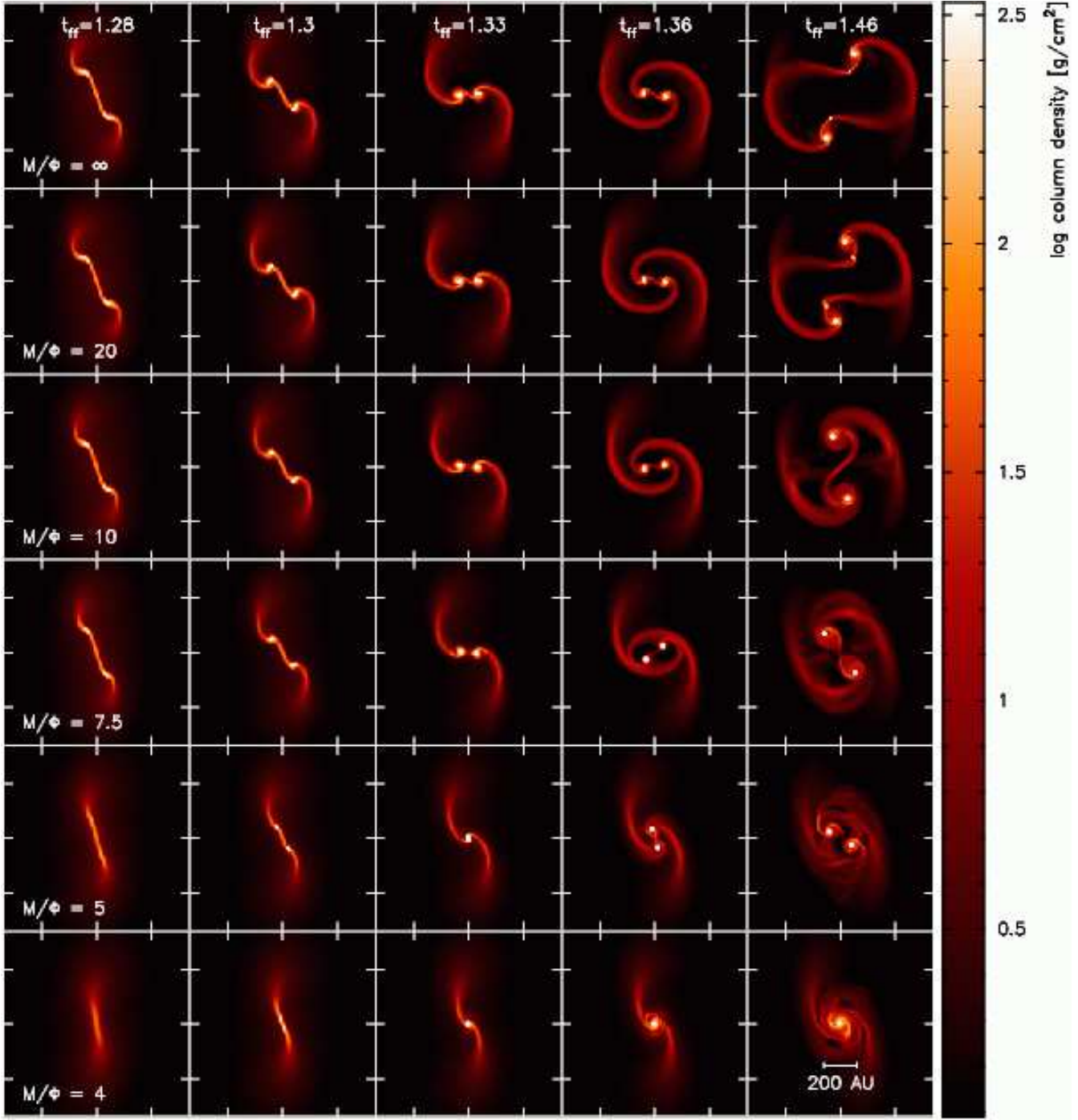,width=\textwidth}
\caption{Results of the binary star formation calculations using a magnetic field aligned with the rotation axis. As in Figure~\ref{fig:axisymcoldens} column density in the fragmenting cloud is plotted, in a sequence of snapshots in time (from left to right) and increasing magnetic field strength (from top to bottom). As previously times are given in units of the free-fall time, $t_{\rm ff}=2.4\times 10^{4}$ yrs and magnetic field strength is expressed in terms of the mass-to-flux ratio in units of the critical value, corresponding to $B=0, 40.7, 81.3, 108.5, 163$ and $203\mu G$ from top to bottom respectively. The separation of the binary system decreases with increasing magnetic field strength, eventually forming only a single star surrounded by a rotationally-dominated circumstellar disc.}
\label{fig:binarycoldens}
\end{center}
\end{figure*}
 
  The initial density perturbations with amplitudes $A=0.1, 0.2$ were applied, retaining equal particle masses, by perturbing the particle positions from a uniform distribution according to the linearized continuity equation
\begin{equation}
\nabla\cdot(\delta {\bf r}) = \frac{\delta \rho}{\rho_{0}}.
\end{equation}
Using (\ref{eq:rhoperturb}) the perturbation in position is given by
\begin{equation}
\delta \phi = -\frac{A}{2} \sin{(2\phi_{0})},
\end{equation}
where $\phi_{0}$ is the unperturbed azimuthal angle. For calculations using $A>0.5$ this is no longer a good approximation an we instead apply the perturbation by changing particle masses appropriately.

\subsubsection{Initial field aligned with the rotation axis}
 The results of the binary star formation calculations with the initial magnetic field aligned with the rotation axis are shown in Figure~\ref{fig:binarycoldens}, where as previously we have computed a series of runs of increasing magnetic field strength, corresponding to mass-to-flux ratios in units of the critical value of (from top to bottom) $\infty$ (that is, hydrodynamics), $20, 10, 7.5, 5$ and $4$. Given the initial temperature and density of the cloud, the corresponding values for the ratio of gas to magnetic pressure are $\beta_{\rm m} = \infty, 39, 9.8, 5.5, 2.4$ and $1.57$ respectively. The results are shown at 5 different times (left to right) given in units of the free-fall time ($2.4\times 10^{4}$yrs). The global evolution of the cloud prior to star formation is similar to that discussed in the axisymmetric case (see \S\ref{sec:axisym}, above). Collapse in the runs with $M/\Phi \leq 3$ is strongly suppressed by the magnetic field and these runs are therefore not shown. 
 
 In the binary case there is a clear trend of delayed collapse and decreasing binary separation as the magnetic field strength increases (top to bottom in Figure~\ref{fig:binarycoldens}). Eventually, for very strong magnetic fields ($M/\Phi \leq 4$) binary formation is inhibited completely (the remnant of the initial $m=2$ density perturbation is just visible at $t_{\rm ff} = 1.3$ in the $M/\Phi = 4$ case, but merges to form a single protostar at $t_{\rm ff}=1.33$ and is subsequently surrounded by a massive, circumstellar disc). We follow the calculations to the point at which sub-fragmentation occurs in the weak magnetic field runs (ie. at $t_{\rm ff}=1.46$ for the hydrodynamic and $M/\Phi = 20$ runs).
 
\begin{figure}
\begin{center}
\epsfig{file=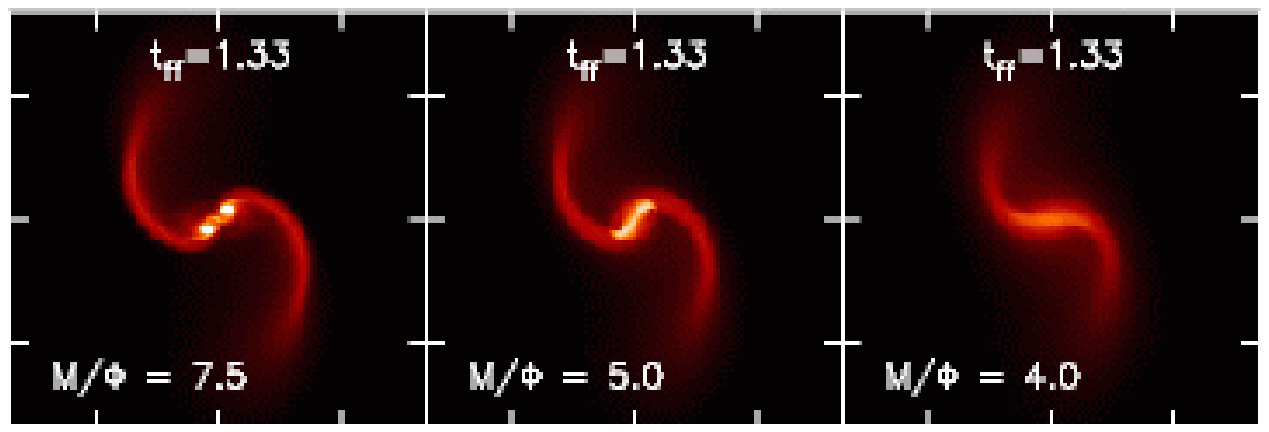,width=\columnwidth}
\vspace{2mm}
\epsfig{file=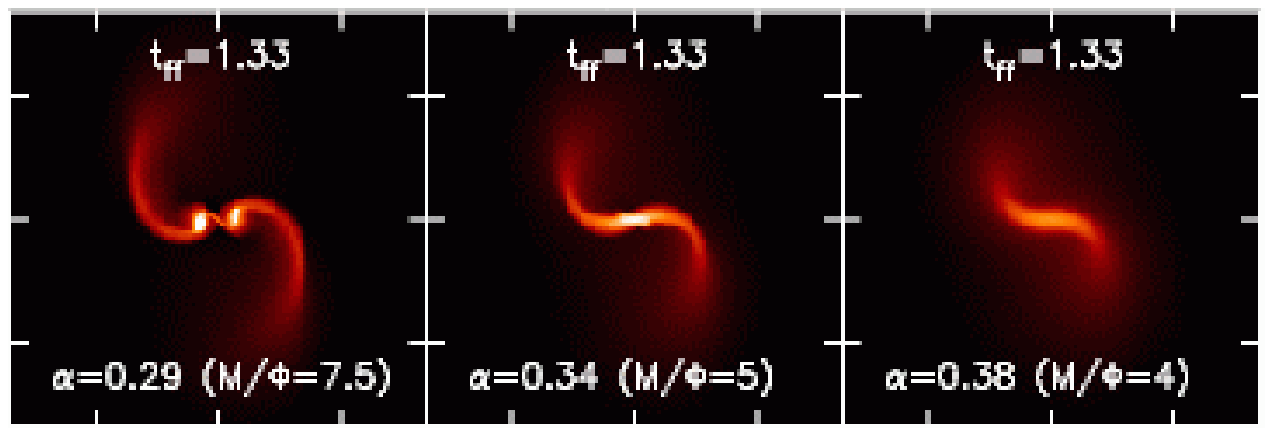,width=\columnwidth}
\caption{Results of (top row) simulations using a magnetic field initially aligned with the rotation axis but without magnetic tension forces and (bottom) purely hydrodynamical simulations using ratios of thermal to gravitational energy, $\alpha$ equivalent to the effective support provided by both the thermal and magnetic pressures. The plots are shown at $t_{\rm ff} = 1.33$ for simulations with mass to flux ratios (top row) of (from left to right) 7.5, 5 and 4, corresponding to the centre panel of the last three rows of Figure~\ref{fig:binarycoldens}. In the hydrodynamical case (bottom row) we have used $\alpha = 0.29, 0.34$ and $0.38$ respectively. In both cases the same transition from a binary to a single protostar is observed, indicating that, for fields aligned with the rotation axis, magnetic pressure plays the dominant role in suppressing binary formation and that this is similar to an increase in the effective thermal energy of the cores.}
\label{fig:binarynotens}
\end{center}
\end{figure}

 Whilst it is tempting to attribute the decrease in binary separations observed in Figure~\ref{fig:binarycoldens} to angular momentum transport induced by the magnetic field (ie. via magnetic braking of the core), this is in fact \emph{not} the dominant cause. Rather, the magnetic pressure acts to increase the effective ``$\alpha$'' (ratio of thermal to gravitational energy) in the cloud which reduces the propensity of the cloud to fragment. That this is the case is demonstrated in Figure~\ref{fig:binarynotens} where we show the results of three runs with $M/\Phi=7.5, 5$ and 4 that were performed a) without magnetic tension forces (top row of Figure~\ref{fig:binarynotens}) and b) with no magnetic fields, but with the cloud temperature increased to give equivalent effective values of $\alpha$ (where the corresponding values are $\alpha=0.29,0.34$ and $0.38$ respectively, determined by transferring the initial magnetic energy to the thermal energy of the cloud). The results are shown at $t_{\rm ff} = 1.33$, corresponding to the centre panels of the fourth, fifth and sixth rows in Figure~\ref{fig:binarycoldens}. The interesting point to note is that the trend in binary separation is similar in \emph{both} cases (there is a transition from a binary to a single protostar), indicating that it is the magnetic pressure which is providing the dominant effect in suppressing binary formation and that the effect of magnetic pressure is similar to an increase in the effective thermal energy of the cloud. That is not to say that magnetic tension forces are without effect as there are differences between the results with and without magnetic tension forces. Nor is the effect of magnetic pressure \emph{exactly} equivalent to a thermal pressure, indicated by the slight differences which remain between the hydrodynamical calculations and the magnetic pressure-only calculations and the dependence on field geometry (see \S\ref{sec:binaryBx}, below). However, the dominant effect observed in Figure~\ref{fig:binarycoldens} is primarily attributable to magnetic pressure effects rather than magnetic braking.

\begin{figure}
\begin{center}
\begin{turn}{270}\epsfig{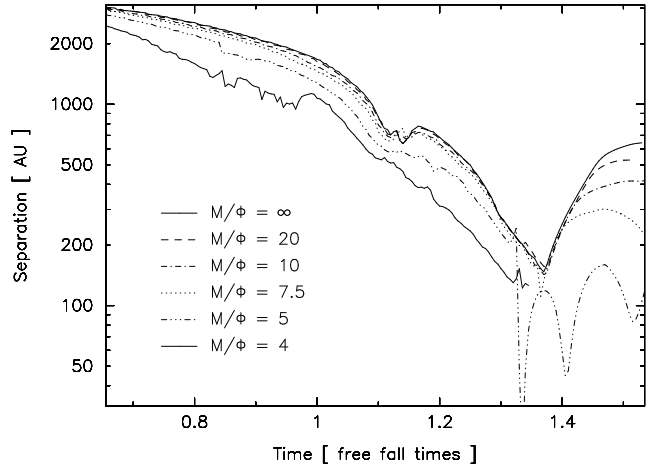}\end{turn}
\caption{Separation of the binary systems shown in Figure~\ref{fig:binarycoldens} plotted as a function of time, with field strengths (mass to flux ratios) indicated by the legend. Prior to sink particle formation, we define the binary separation as the distance between the two highest density maxima constrained to be in opposite hemispheres. The plot quantifies the decrease in binary separation with increasing magnetic field strength seen in Figure~\ref{fig:binarycoldens}. The dip at $t_{\rm ff}\sim 1.1$ in the low magnetic field strength runs is a spurious feature due to a transient density peak related to particle gridding effects.}
\label{fig:sep}
\end{center}
\end{figure}

 The results shown in Figure~\ref{fig:binarycoldens} are quantified in Figure~\ref{fig:sep} which shows the binary separation as a function of time for the magnetic field strengths shown in Figure~\ref{fig:binarycoldens}. Prior to sink particle formation, we define the binary separation as the distance between the two highest density maxima constrained to be in opposite hemispheres. Some spurious effects from this definition are visible at $t_{\rm ff}\sim 1.1$ in the weak magnetic field runs, where there is a transient density maximum due to particle gridding effects. Also the results become meaningless when only one star is formed, as in the $M/\Phi=4$ case beyond $t_{\rm ff}\sim 1.3$ (lowest solid line) at which point we do not plot any more points. However, the general trend is clear - namely that the binary separation decreases monotonically as the magnetic field strength is increased. The $M/\Phi=5$ case can be seen to form a very close but eccentric binary system which goes through a closest approach of $\sim 30$ AU.
 
 \begin{figure*}
\begin{center}
\epsfig{file=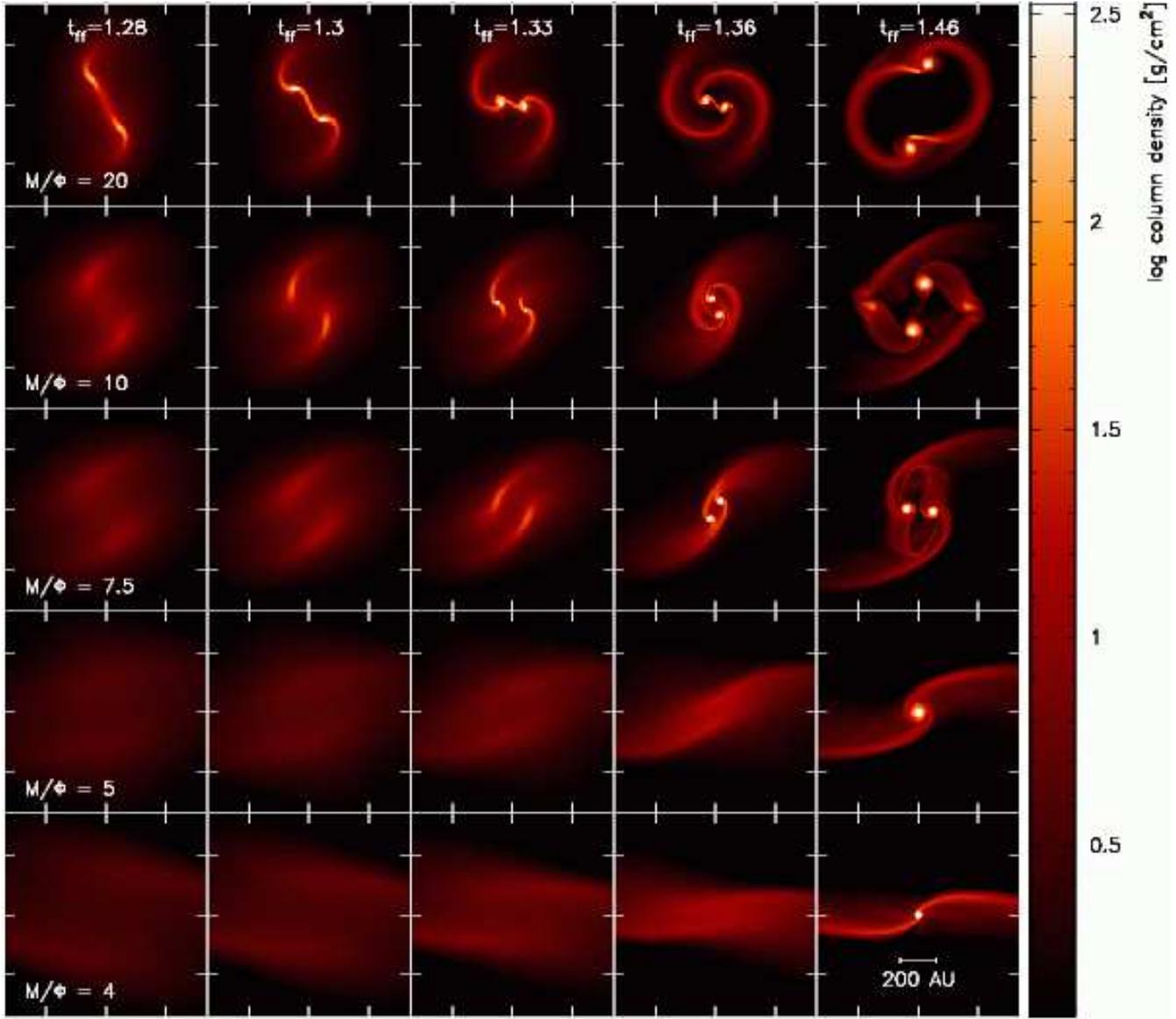,width=\textwidth}
\caption{Results of binary star formation calculations using a magnetic field initially oriented perpendicular to the rotation axis (ie. initial field in the $x-$direction). As previously times are given in units of the free-fall time, $t_{\rm ff}=2.4\times 10^{4}$ yrs and magnetic field strength is expressed in terms of the mass-to-flux ratio in units of the critical value, corresponding to $B=40.7, 81.3, 108.5, 163$ and $203\mu G$ from top to bottom respectively.  The transition from a binary to a single protostar occurs at a lower field strength than with the initial field aligned with the rotation axis.}
\label{fig:binarycoldensBx}
\end{center}
\end{figure*}

\begin{figure*}
\begin{center}
\epsfig{file=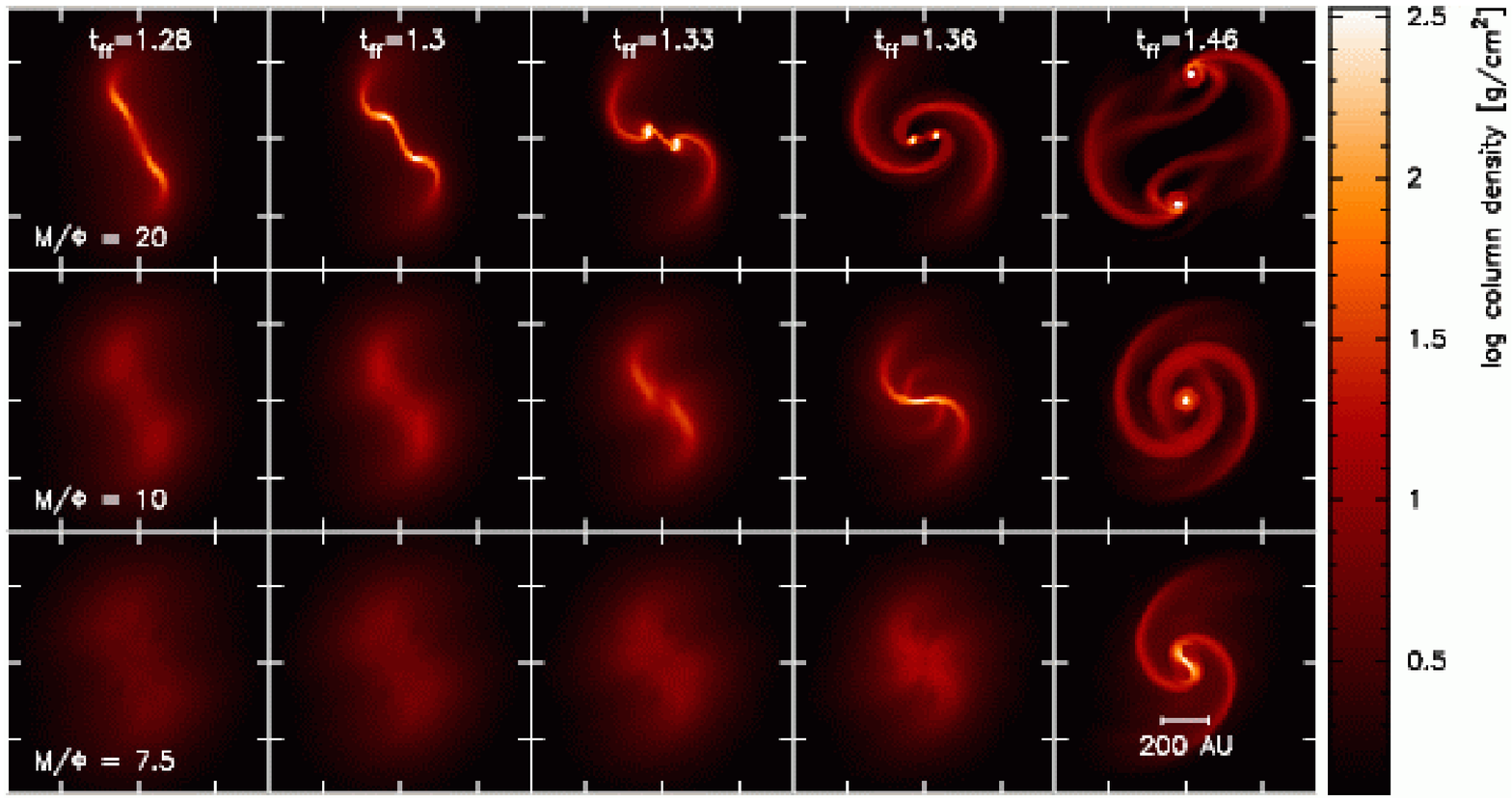,width=\textwidth}
\caption{Results of binary star formation calculations using a magnetic field initially oriented perpendicular to the rotation axis (ie. initial field in the $x-$direction), as in Figure~\ref{fig:binarycoldensBx} but with magnetic tension forces turned off.  The transition from a binary to a single protostar occurs at a {\it lower} field strength when magnetic tension forces are excluded (i.e., magnetic tension aids binary formation).}
\label{fig:binarycoldensBxnotens}
\end{center}
\end{figure*}
 
 For the binary star formation case we have also performed simulations which start from larger initial perturbations ($A=0.2$ and $A=0.5$ percent) and with larger and smaller cloud radii ($5\times 10^{16}$cm and $3\times 10^{16}$cm). The runs with larger initial perturbations are progressively less influenced by the magnetic field. Indeed for the $A=0.5$ percent perturbation all of the calculations up to a mass to flux ratio of $M/\Phi \approx 2$ formed binaries with separations that are not significantly influenced by the magnetic field strength. 
The effect of changing the cloud radius is to change at what stage during the collapse the cloud becomes optically thick (ie. the equation of state changes from isothermal to non-isothermal). Thus disc fragmentation is (relative to the calculations presented here) suppressed for smaller cloud radii and enhanced for larger radii but with similar trends in the influence of the magnetic field.

\subsubsection{Initial field perpendicular to the rotation axis}
\label{sec:binaryBx}
 Results of binary star formation calculations beginning with a magnetic field oriented perpendicular to the rotation axis (that is, with a field initially in the $x-$direction) are shown in Figure~\ref{fig:binarycoldensBx}. As in Figure~\ref{fig:binarycoldens} some general trends are clear: increasing the magnetic field strength leads to a delayed collapse and increasingly suppresses binary formation. In this case, however, the transition from a binary to a single star occurs earlier (that is, a single star is formed at $M/\Phi = 5$ in Figure~\ref{fig:binarycoldensBx} compared to $M/\Phi=4$ in Figure~\ref{fig:binarycoldens}) and the binary perturbation is increasingly deformed by the magnetic field, which at higher field strengths results in a ``double bar-like collapse'' (most evident in the higher field strength runs in Figure~\ref{fig:binarycoldensBx}. 
 
 As previously, the global trends (delayed collapse and transition to a single star) are the result of the extra support provided to the cloud by magnetic pressure alone. This is demonstrated by Figure~\ref{fig:binarycoldensBxnotens} which shows the results of similar calculations (that is, with fields initially perpendicular to the rotation axis) but with magnetic tension forces turned off.  In this case the transition to a single star occurs for even lower magnetic field strengths (at $M/\Phi = 10$). This indicates not only that magnetic pressure is providing the dominant role in suppressing fragmentation but also that magnetic tension can act to dilute the effect of magnetic pressure, even \emph{aiding} binary formation. We note that \citet{boss00,boss02} similarly concluded that magnetic tension forces can act to promote fragmentation, albeit using a simplistic approximation to model the affect of magnetic fields.

 Comparing the middle column of Figure~\ref{fig:binarycoldensBxnotens} to the pressure-only calculations with an aligned magnetic field shown in Figure~\ref{fig:binarynotens} also demonstrates that the effect of magnetic pressure is dependent on the field geometry, acting more like a equivalent thermal pressure when the field is aligned with the rotation axis.
 
\begin{figure*}
\begin{flushleft}
\begin{turn}{270}\epsfig{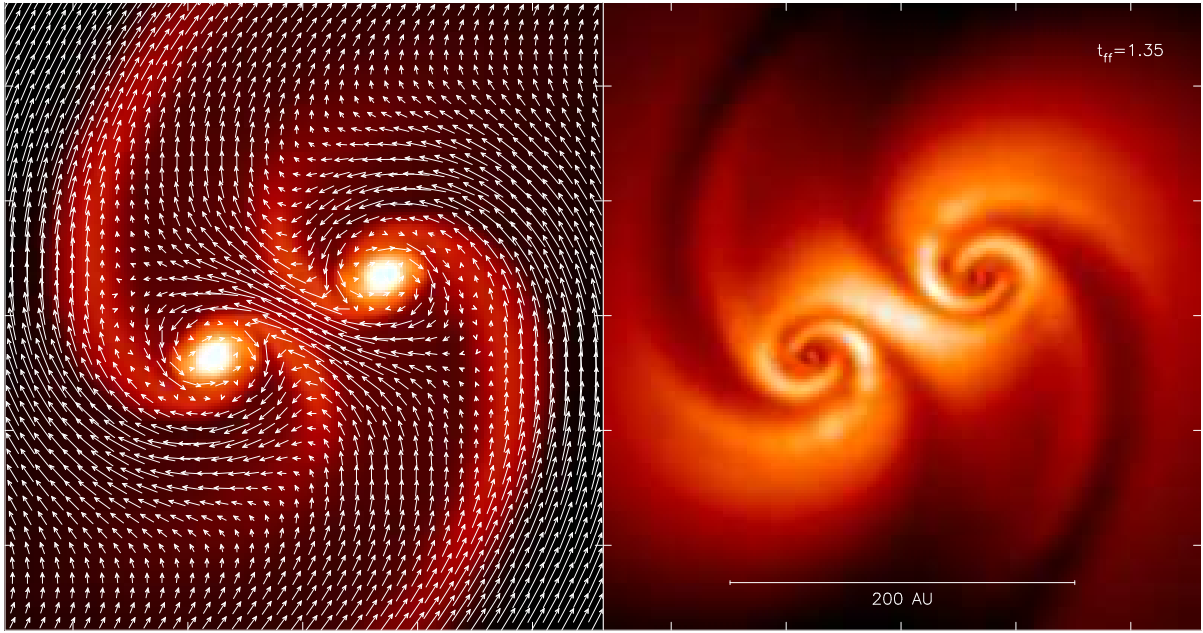}\end{turn}
\caption{Magnetic cushioning in action. The panels show column density and integrated magnetic field vectors (left) and integrated magnetic pressure (right) at $t_{\rm ff} = 1.35$ in the $M/\Phi = 10$ run (corresponding to the second row of Figure~\ref{fig:binarycoldensBx}). The magnetic field, initially in the orbital plane, is wound up by the differentially rotating cloud to form a ``cushion'' between the binary, preventing it merging into a single protostar.  Thus, the magnetic cushion aids binary formation.}
\label{fig:magcushion}
\end{flushleft}
\end{figure*}

  The deformation of the binary perturbation in Figure~\ref{fig:binarycoldensBx} is not evident in the tension-free calculations (Figure~\ref{fig:binarycoldensBxnotens}), indicating (as one might expect) that this effect is the result of the gas being squeezed by the magnetic field lines. It is this squeezing due to magnetic tension that acts to hold up the rapid transition to a single star with increasing field strength observed in the tension-free runs and thus dilute the effect of magnetic pressure in suppressing fragmentation. The magnetic field, being in this case aligned \emph{along} the binary perturbation, effectively acts as a ``cushion'' between the two stars which prevents their merging. This effect, which we henceforth refer to as ``magnetic cushioning'', is graphically illustrated in Figure~\ref{fig:magcushion} which shows the magnetic field (arrows in left panel, overlaid on a column density plot) and integrated magnetic pressure (right panel) in the $M/\Phi = 10$ run (corresponding to the second row of Figure~\ref{fig:binarycoldensBx}) at $t_{\rm ff}=1.35$. The ``cushion'' formed by the magnetic field between the two stars is clearly evident, and it is this ``magnetic cushion'' which prevents the binary system from merging to form a single star (and also produces the wonderful symmetry in the spiral arms). 

\begin{figure}
\begin{center}
\begin{turn}{270}\epsfig{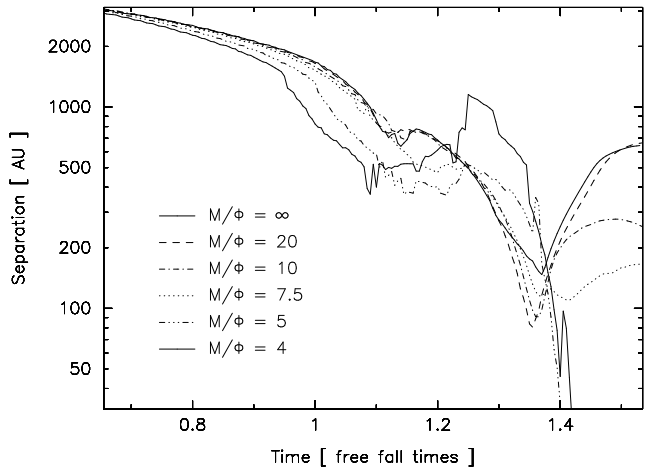}\end{turn}
\caption{Separation of the binary systems shown in Figure~\ref{fig:binarycoldensBx} plotted as a function of time, with field strengths (mass to flux ratios) indicated by the legend. }
\label{fig:sepBx}
\end{center}
\end{figure}

 The results shown in Figure~\ref{fig:binarycoldensBx} are quantified in Figure~\ref{fig:sepBx} which shows the binary separation as a function of time for the magnetic field strengths shown in Figure~\ref{fig:binarycoldensBx} and may be compared with the corresponding figure (Figure~\ref{fig:sep}) for the runs with the field aligned with the rotation axis. As previously, prior to sink particle formation, we compute the separation of two density maxima in opposite hemispheres. In the stronger field runs ($M/\Phi = 4$ and $M/\Phi=5$) the binary perturbation is strongly deformed by the magnetic field, producing the observed increase in separation observed at  $t_{\rm ff}\sim 1.2$. The binary separations are in each case smaller than the equivalent runs using a field aligned with the rotation axis, which demonstrates that the effect of the magnetic field of the binary system is stronger in this case. The effect of magnetic cushioning is also apparent in the fact that the runs with $M/\Phi = 20, 10$ and $7.5$ show a trend of \emph{increasing} binary separation at closest approach ($t_{\rm ff}\sim 1.35$), in contrast to Figure~\ref{fig:sep} (although all the separations are smaller than in the aligned-field runs).
 
\section{Discussion}
\label{sec:discussion}

We have conducted a study of how magnetic fields affect the collapse of homogenous molecular cloud cores and cores with initial $m=2$ density perturbations.  In both cases, the 
presence of a magnetic field produces a delayed collapsed, with a longer delay for stronger fields.  This affect is easily attributed to the effect of the magnetic pressure on the collapse.  The magnetic field gives
extra support to the cloud over the thermal pressure alone; rather than acting like a cloud whose
ratio of thermal energy to the magnitude of gravitational energy $\alpha=0.26$ (or $0.35$ in the axisymmetric models), the effect of the
magnetic field is to raise the effective value of $\alpha$.

\subsection{The effect of magnetic fields on protostellar discs}

In the homogenous simulations, we find that a single protostar (sink particle) is formed and surrounded by a disc.  Stronger magnetic fields lead to a delay in the formation time of the protostar as mentioned above, but they also decrease the rate of accretion onto the disc.  The disc radius also increases more slowly with time.  It is well known that the rate of infall of mass onto a massive disc is crucial in generating gravitational instabilities (e.g. \citealt{bonnell94,whitworth95,hennebelle04}) and, indeed, we see this effect here.  In the purely hydrodynamical case, the disc surrounding the protostar is gravitationally unstable and exhibits strong spiral density waves soon after the protostar forms (although the instability is not strong enough to force the disc to fragment).  With a magnetic field initially aligned with the rotation axis, the slower rate of mass infall onto the disc leads to a weakening of the gravitational instability such that for mass to flux ratios less than $M/\Phi \approx 10$ the spiral features are very weak (Figure \ref{fig:axisymcoldens}).  With a field initially perpendicular to the rotation axis, the gravitational instability is very weak even for $M/\Phi = 20$ (Figure \ref{fig:axisymcoldens_Bx}).

Gravitational instabilities in protostellar discs may be important for several reasons.  First, if the gravitational instability is strong enough, the disc may fragment to form a companion (e.g. \citealt{bonnell94,bb94a,bb94b,whitworth95,rice05}).  This is particularly relevant to the magnetised star formation simulations performed by \citet{hw04b}.  They began with a rotating cloud that, in the absence of magnetic fields formed a single object surrounded by a gravitationally unstable disc that fragmented to form companions.  With magnetic fields initially aligned with the rotation axis, they found that the disc was much smaller and did not fragment. This is consistent with our simulations in that we also find that magnetic fields reduce the tendency for a disc to be gravitational unstable.  However, Hosking \& Whitworth attributed the inhibiting of fragmentation to the loss of angular momentum due to magnetic tension forces.  Here we find that the effect of magnetic pressure in decreasing the mass infall rate on to the disc may be just as important in suppressing disc fragmentation.

Second, if a protostellar disc is gravitationally unstable (but not so strongly as to fragment), spiral density waves are likely to be the dominant angular momentum transport mechanism within the accretion disc \citep{lr04,lr05,fromang04}.  It is only later when the disc becomes more stable that the magnetorotational instability is likely to take over as the dominant angular momentum transport mechanism.  Our findings suggest that for magnetised cores the phase of rapid angular momentum transport due to gravitational instabilities will be less important than for stars formed in unmagnetised cores.

Third, it has been suggested that the spiral density waves in gravitationally unstable protoplanetary discs might aid the formation of large (kilometre-sized) planetesimals, and thus planets, by collecting centimetre and metre-sized planetesimals in the spiral waves until the density of solids along the arms is sufficient for gravitational instability to form solid kilometre-sized objects \citep{rice05}.  However, the effect that we observe here of magnetic fields slowing the infall of gas onto the disc and therefore weakening the gravitational instability argues that these processes may be less important for stars forming in magnetised molecular cloud cores.  On the other hand, it is plausible that the spiral structure generated by the magnetic field in the outer parts of the discs formed in the misaligned field calculations may act in a similar manner.

\subsection{The effect of magnetic fields on fragmentation}

Several past studies have cast doubt on the ability of strongly magnetised molecular cloud cores to form anything other than single objects \citep{phillips86a,phillips86b,hw04b}.  Even MHD calculations that have obtained binary systems have either produced binaries with very small separations \citep{bp06} or have required strongly magnetised cores to have very large initial rotation rates \citep{machida05b}.  Given that binary and multiple stellar systems are very common, this is potentially a serious problem.  

Perhaps the easiest way to obtain a binary system is to begin with a strongly perturbed initial core -- a picture that is also in qualitative agreement with observations \citep{pringle89}.  Therefore, we decided to start with strongly perturbed cores and examine the effect of magnetic fields in such cases.  We performed simulations of molecular cloud cores with initial $m=2$ density perturbations with amplitudes of 10, 20 and 50 percent.  Initial conditions with 50 percent density perturbations showed essentially no dependence of the results on the magnetic field up to mass-to-flux ratios of $M/\Phi \approx 2$ for a field that was initially aligned with the rotation axis -- all the calculations formed binaries and their separations were essentially indistinguishable.  Even with a magnetic field initially perpendicular to the rotation axis, although there was a decrease of the binary's separation with the initial field strength, binaries were formed up to mass-to-flux ratios of $M/\Phi \approx 3$. This demonstrates that {\it if the initial density perturbations are large enough, even strong magnetic fields are not able to suppress or even significantly alter the pattern of fragmentation}.  We note that this result is in qualitative agreement with the single MHD calculation of binary formation performed by \citet{ziegler05}.

As presented in Section \ref{sec:binary}, for spherical molecular cloud cores with weaker 10 percent $m=2$ density perturbations we find that the fragmentation does depend on the field strength and the alignment of the field relative to the rotation axis.  A stronger field results in a closer binary with only a single protostar being produced when the the mass-to-flux ratio is reduced to below $(M/\Phi) \approx 4$ or $(M/\Phi) \approx 8$ for initial fields that are aligned or perpendicular to the the rotation axis, respectively.  However, the question
arises as to whether this effect is primarily due to the increased support provided by the magnetic field 
(i.e. an increase in the effective $\alpha$ of the cloud) or due to the loss of angular momentum 
because of the magnetic braking provided by magnetic tension forces.  It is well established
that clouds with higher initial values of $\alpha$ are less prone to fragmentation (e.g. \citealt{mhn84,ti99}).  As we have demonstrated above, re-running the $m=2$ calculations in which the field is initially aligned with the rotation axis without magnetic tension, or indeed without magnetic fields at all but with a thermal $\alpha$ (i.e.\ increased sound speed) that is set equal to the effective $\alpha$ in the MHD calculations, gives very similar results to using the full MHD equations.  In particular, we find that the trend of decreasing separation followed by the transition to a single 
protostar as the magnetic field strength is increased is almost entirely due to the 
extra support provided by the magnetic field.  When the magnetic field and the rotation axis are initially perpendicular, the transition to a single protostar occurs at lower field strengths than with an aligned field.  However, in these cases {\it omitting} magnetic tension forces results in even weaker field strengths producing single protostars.  Unexpectedly, magnetic tension {\it aids} fragmentation in these cases due to magnetic cushioning between the two initial density perturbations.  In summary, previous studies of the effect of magnetic fields on fragmentation have generally attributed the suppression of fragmentation to magnetic braking via magnetic tension/torsion forces.  However, our results show that it is not this straightforward.  In the calculations presented here, {\it magnetic pressure plays the dominant role in suppressing fragmentation, while magnetic tension either has little effect or aids fragmentation, depending on the field geometry.}

\section{Summary}
\label{sec:summary}

We have performed magnetohydrodynamic (MHD) simulations of the collapse of molecular cloud cores, some with initial $m=2$ density perturbations, and others that were homogenous.  

In terms of the effect of magnetic fields on the fragmentation of perturbed molecular clouds, we have two main conclusions.  First, we find that wide binaries can be readily obtained from perturbed molecular cloud cores even with mass-to-flux ratios as low as $M/\Phi = 2-3$.  Since most molecular cloud cores are observed to be strongly aspherical, {\it we conclude that magnetic fields may not be as significant a problem to binary formation as many past studies have suggested}.  Second, in agreement with past studies, we find that magnetic fields act to suppress fragmentation.  However, contrary to past studies that have emphasized the importance of magnetic tension forces and magnetic braking in suppressing fragmentation, we find that {\it the extra support given by magnetic pressure over the thermal support is the dominant reason for the suppression of fragmentation and that magnetic tension can actually aid fragmentation}.

For all calculations, we find that stronger magnetic fields result in longer delays to the collapse due to the increased support provided by the field above that provided by thermal pressure alone.  The delaying of the collapse has a potentially crucial effect on protostellar discs that form around protostars.  Because the infall rate of mass onto the disc from the envelope is reduced, the discs are less prone to gravitational instabilities.  This may have at least three important effects.  First, high infall rates on to discs have been shown in the past to be responsible for driving gravitational instabilities that are strong enough to result in disc fragmentation and the production of binary and multiple systems.  Thus, we conclude that magnetic fields tend to inhibit disc fragmentation.  Second, even for initial conditions that in the absence of magnetic fields give infall rates insufficient to cause disc fragmentation, the discs frequently generate spiral density waves that efficiently transport angular momentum.  With magnetic fields, the lowering of the infall rate on to the disc decreases the strength and importance of these instabilities.  Third, spiral density waves have been suggested as a possible mechanism for producing high enough concentrations of metre-sized planetesimals to produce kilometre-sized planetesimals via gravitational instability.  Since we find that magnetic fields decrease the strength of the spiral density waves in protostellar discs, magnetic fields may also decrease the likelihood of forming planetary systems by such a mechanism.

\section*{Acknowledgements} 
We thank the referee, Ant Whitworth, for asking questions that led to a significant improvement in the paper.
DJP is supported by a UK PPARC postdoctoral research fellowship.  MRB is grateful for the support of a 
Philip Leverhulme Prize and a EURYI Award. Calculations were performed using the School of Physics iMac cluster at the University of Exeter and on the United Kingdom Astrophysical Fluids Facility (UKAFF). We thank Charles Williams in particular for support on the iMac cluster. Visualisations made use of SPLASH/SUPERSPHPLOT, a visualisation tool for SPH that is publicly available at http://www.astro.ex.ac.uk/people/dprice/splash.  This work, conducted as part of the award ``The formation of stars and planets: Radiation hydrodynamical and magnetohydrodynamical simulations"  made under the European Heads of Research Councils and European Science Foundation EURYI (European Young Investigator) Awards scheme, was supported by funds from the Participating Organisations of EURYI and the EC Sixth Framework Programme.

\appendix

\bibliography{sph,mhd,starformation}

\label{lastpage}
\enddocument